\documentclass[prl,floatfix,shwopacs,amsmath,twocolumn]{revtex4}
\usepackage{amssymb}
\usepackage{graphicx}

\begin{document}

\newcommand{\one}{\ensuremath{\hbox{$\mit I$\kern-.3em$\mit I$}}}
\renewcommand{\one}{\ensuremath{\hbox{$\mathrm I$\kern-.6em$\mit 1$}}}
\renewcommand{\one}{\ensuremath{\hbox{$\mathrm I$\kern-.6em$\mathrm 1$}}}
\newcommand{\ad}{\ensuremath{a^\dagger}}
\newcommand{\bd}{\ensuremath{b^\dagger}}
\newcommand{\cd}{\ensuremath{c^\dagger}}
\newcommand{\dd}{\ensuremath{d^\dagger}}
\newcommand{\ed}{\ensuremath{e^\dagger}}
\newcommand{\ket}[1]{\ensuremath{|{#1}\rangle}}
\newcommand{\bra}[1]{\ensuremath{\langle{#1}|}}
\newcommand{\braket}[1]{\ensuremath{\langle{#1}\rangle}}
\newcommand{\be}{\begin{equation}}
\newcommand{\ee}{\end{equation}}
\newcommand{\bea}{\begin{eqnarray}}
\newcommand{\eea}{\end{eqnarray}}
\newcommand{\bma}{\begin{mathletters}}
\newcommand{\ema}{\end{mathletters}}
\def\<{\langle}
\def\>{\rangle}

\title{Trapping atoms in the vacuum field of a cavity}

\author{C. \surname{Sch\"on}}
\email{Christian.Schoen@mpq.mpg.de}
\affiliation{Max-Planck-Institut f\"ur Quantenoptik,
  Hans-Kopfermann-Str. 1, Garching, D-85748, Germany.}
\author{J.~I. \surname{Cirac}}
\affiliation{Max-Planck-Institut f\"ur Quantenoptik,
  Hans-Kopfermann-Str. 1, Garching, D-85748, Germany.}

\date{\today}

\begin{abstract}
The aim of this work is to find ways to trap an atom in a cavity.
In contrast to other approaches we propose a method where the
cavity is basically in the vacuum state and the atom in the ground
state. The idea is to induce a spatial dependent AC Stark shift by
irradiating the atom with a weak laser field, so that the atom
experiences a trapping force. The main feature of our setup is
that dissipation can be strongly suppressed. We estimate the
lifetime of the atom as well as the trapping potential parameters
and compare our estimations with numerical simulations.
\end{abstract}

\maketitle

%%%%%%%%%%%%%%%%%%%%%%%%%%%%%%%%%%%%%%%%%%%%%%%%%%%%%%%%%%%%%
\section{Introduction}

Cavity QED constitutes one of the most important experimental
set--ups where the basic properties of Quantum Mechanics can be
controlled, observed, and tested. During the last years, a
significant experimental progress has taken place, allowing to
observe quantum phenomena in the interaction of a single atom with
a single mode of the electromagnetic field, both in the optical
\cite{Kimble00heat,Kimble00cool,
Rempe99obs,Rempe02kal,Rempe02src,Feld98,Walther01opt,Blatt02} and
microwave regime \cite{Walther01mic,Haroche02}. Some of these
experiments are currently limited by the fact that (neutral) atoms
typically move almost freely in the cavity and eventually leave
it, which restricts the duration of the experiment as well as its
controllability. For example, in the optical regime, the coupling
between the atoms and the cavity mode strongly depends on the
position of the atom, and thus when it moves this can strongly
affect the interaction.

In order to overcome these problems, several strategies to trap an
atom in a cavity have been put forward
\cite{Haroche91trap,Ritsch97,Ritsch98,Kimble99trap,Kimble00trap,Kimble01trap,
Rempe00trap,Rempe00trap_long,Kimble99fort,Kimble02fort,Chapman01}.
Some of them involve using some external laser fields which exert
a confining force to the atom, something that has been
successfully realized in recent experiments
\cite{Kimble99fort,Kimble02fort,Chapman01}. In a far-off resonant
trap (FORT) this is achieved by employing a far-off resonant
trapping beam along the cavity axis. A more intriguing approach
consists of using the cavity mode itself to confine the atom
\cite{Haroche91trap,Kimble01trap,Ritsch97,Ritsch98}. In remarkable
experiments
\cite{Kimble99trap,Kimble00trap,Rempe00trap,Rempe00trap_long} it
has been possible to keep an atom in a cavity just using the force
provided by a single photon. In this work we will show that it is,
in principle, possible to trap an atom in its ground state in a
cavity which is basically in the vacuum state. Apart from its
fundamental interest, our method may have some practical
advantages with respect to the previous one in that since the atom
and the cavity mode are in their ground state, losses are
appreciable reduced.

Atom trapping in cavities has interesting applications in the
field of quantum information. All the proposals of quantum
computation using atoms interacting via a common cavity mode
\cite{Pellizzari95} require that the atoms are trapped in the
cavity in a fixed position. Moreover, this is also required to
build quantum networks involving cavity QED setups
\cite{Cirac97,vanEnk98,Pellizzari97}. In those cases the idea is
to store the quantum information in two internal ground levels of
each atom, $|g\rangle$ and $|g'\rangle$, and to entangle them by
using real or virtual photon exchange through the cavity mode.
Note that in this context a single spontaneous emission or cavity
loss may have dramatic effects for all quantum information tasks
(see, however, Ref.\ \cite{vanEnk97}). Thus, it is not only
important to trap the atoms in the cavity but also to reduce the
decoherence time as much as possible. By trapping the atoms in the
vacuum state of the cavity our scheme achieves a strong reduction
of the decoherence processes.

The plan of this paper is as follows. In Section II we give a
qualitative description of our scheme, estimating its operating
conditions such as the depth of the trapping potential and the
lifetime of the state. In Section III we give a full description
of the method including dissipative processes. The analytical
results and estimations are checked numerically in Section IV.
Finally, Section V contains a summary of our results.

%%%%%%%%%%%%%%%%%%%%%%%%%%%%%%%%%%%%%%%%%%%%%%%%%%%%%%%%%%%%%
\section{Description}

\begin{figure}[ht]
\includegraphics[width=4cm]{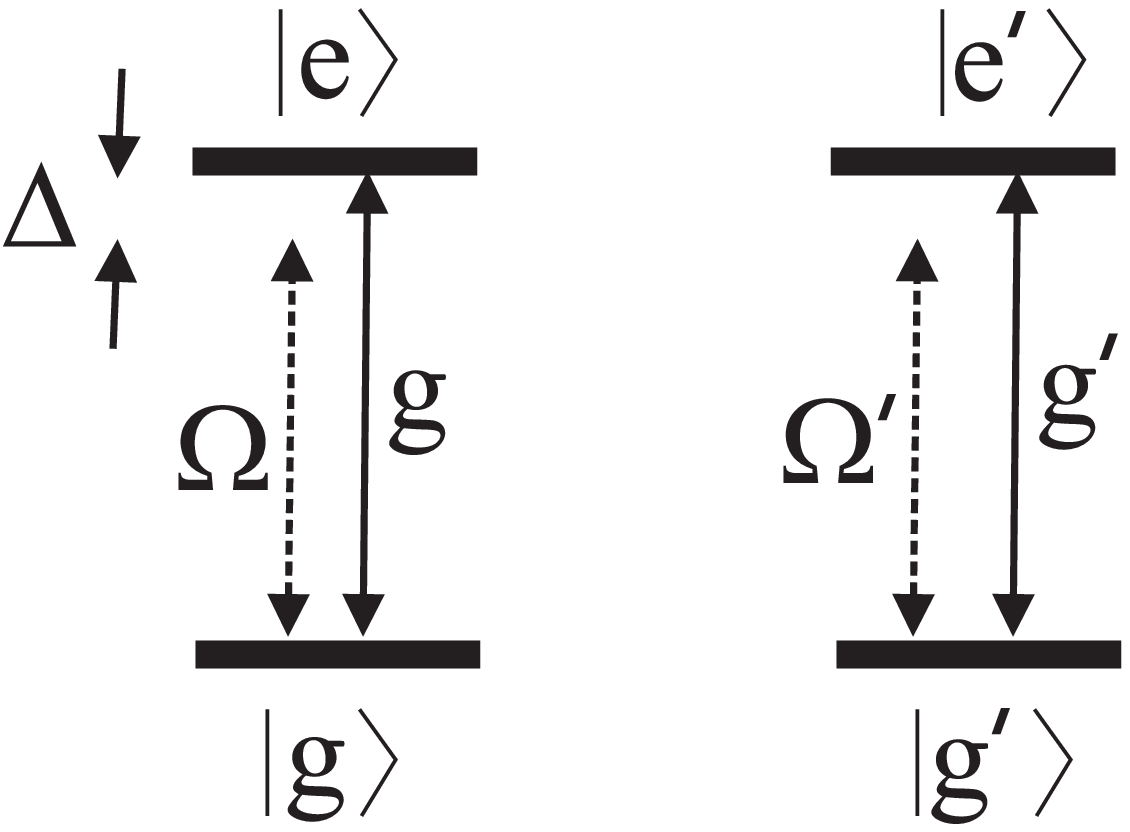}
\caption{Level scheme of the atom. The two ground levels
$|g\rangle$ and $|g'\rangle$ are resonantly coupled by two cavity
modes to two excited levels $|e\rangle$ and $|e'\rangle$. The
coupling strength is given by $g$ and $g'$. There is an additional
external laser field which couples to the atomic transitions with
the Rabi frequencies $\Omega$ and $\Omega'$.} \label{eegg}
\end{figure}

We consider an atom with two internal ground levels $|g\rangle$
and $|g'\rangle$, which are resonantly coupled by two cavity modes
to two excited levels $|e\rangle$ and $|e'\rangle$, respectively.
Additionally, an external plane-wave laser field detuned by
$\Delta$ excites the same transition (see Fig.\ \ref{eegg}).
Note that this laser field does not exert any force on the atoms.
In the following we will consider only the levels $|g\rangle$ and
$|e\rangle$, since for the other two levels the same description
applies (they are independent).

\begin{figure}
\includegraphics[width=8cm]{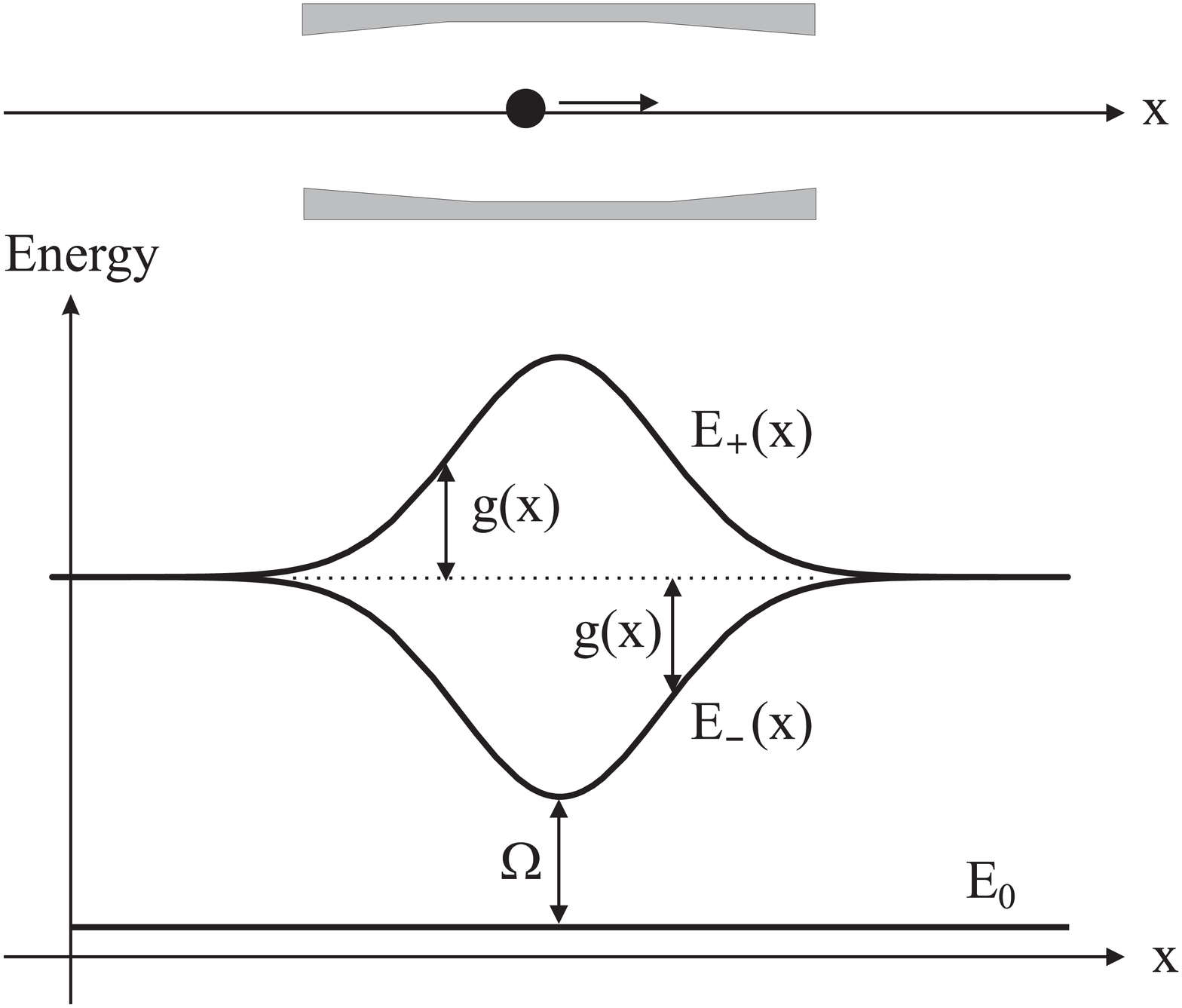}
\caption{Setup and instantaneous energy levels of the atom as a
function of its position. $E_{\pm}(x)$ and $E_0$ are the energies
for the upper and lower dressed state and the ground state. The
external laser field is on resonance with the transition
$|g,0\rangle\to |-\rangle$ near $x=0$ and its Rabi frequency is
denoted by $\Omega$.} \label{jc}
\end{figure}

In order to understand the mechanism that we propose, it is
convenient first to revise the method used in previous experiments
\cite{Kimble99trap,Kimble00trap,Rempe00trap,Rempe00trap_long} to
keep an atom in a cavity. The interaction between the atom and
cavity mode is characterized by a coupling constant $g(x)$, which
depends on the atomic position $x$. In Fig.\ \ref{jc} we show the
set--up, as well as the instantaneous energy levels of an atom as
a function of its position (in one dimension).
The ground state of the composed atom--cavity system is
$|g,0\rangle$, where $|n\rangle$ is the cavity state with $n$
photons (in this case $n=0$). The corresponding energy, $E_0$ is
position independent. The first two excited levels are the dressed
states of the Jaynes--Cummings Hamiltonian \cite{jc63},
$|\pm\rangle=\frac{1}{\sqrt{2}}(|g,1\rangle\pm |e,0\rangle)$, with
corresponding energies $E_{\pm}(x)=\pm g(x)$ in the interaction
picture, where we have taken $\hbar=1$ . As Fig.\ \ref{jc} shows,
the position--dependence of $E_-(x)$ provides the atom with a
confining potential at the center of the cavity. Thus, if the atom
can be prepared in the state $|-\rangle$ with a kinetic energy
smaller than $g_0=g(0)$, it will remain trapped
\cite{Haroche91trap,Kimble01trap,Kimble99trap,Kimble00trap,
Rempe00trap,Rempe00trap_long,Ritsch97,Ritsch98}. As the state
$|-\rangle$ contains a linear combination involving one photon,
one can state that the atom is trapped by a single photon. On the
other hand, the state $|-\rangle$ can be efficiently prepared by
starting in the state $|g,0\rangle$ and tuning the external laser
field to be resonant with the transition $|g,0\rangle\to
|-\rangle$ near $x=0$, as indicated in Fig.\ \ref{jc}
\cite{Parkins93}. Note that in the case studied in many references
\cite{Rempe00trap,Rempe00trap_long,Ritsch97,Ritsch98} the trapping
force may be velocity dependent since they are also interested in
laser cooling, whereas for us this is not the case. In this sense
it will be difficult to exactly compare the results of both
approaches.

The above discussion has omitted an important element which is
present in all experiments, namely the dissipation mechanism. On
the one hand, excited atoms may decay very fast (as long as the
state $|e\rangle$ does not correspond to some Zeeman level, which
is coupled to the cavity mode by some Raman transition
\cite{Kimble99ram}). More importantly, cavities have usually
losses, so that the photons will leave the cavity after some time
$t\simeq 1/\kappa$, where $\kappa$ is the cavity damping rate. Any
of these mechanisms will induce the spontaneous transition
$|-\rangle\to |g,0\rangle$, and therefore the atom will no longer
experience the trapping force. The typical time scale of this
processes is of the order of $\Gamma^{-1}$ and $\kappa^{-1}$,
where $\Gamma$ and $\kappa$ are the spontaneous emission and the
cavity damping rate, respectively. In practice
\cite{Kimble99trap,Kimble00trap,Rempe00trap,Rempe00trap_long} the
atom can be promoted several times to the state $|-\rangle$ by the
external laser, so that the trapping time inside the cavity can be
several hundreds of $\kappa^{-1}$. Note, however, that these
spontaneous transitions will break the atomic coherence if we are
using more internal levels to store, for example, some quantum
information in the atom (see Fig.\ \ref{eegg}).

\begin{figure}
\includegraphics[width=8cm]{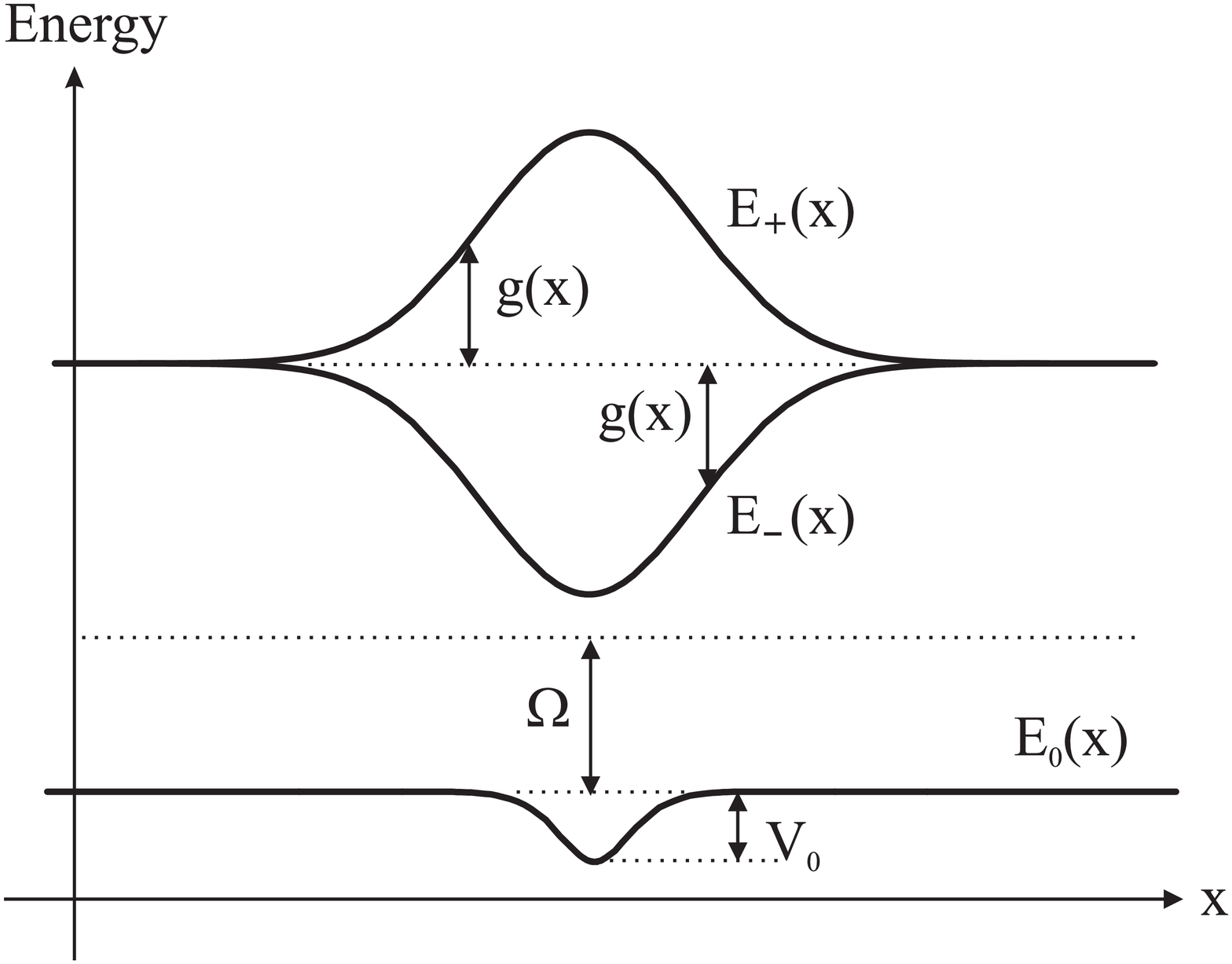}
\caption{The detuned external laser with Rabi frequency $\Omega$
leads to a position dependent AC--Stark shift of $|g,0\rangle$. If
the system is in the level $|g,0\rangle$ its energy $E_0(x)$
depends on the position of the atom. Therefore there exists a
trapping force towards the center of the cavity.} \label{jc_stark}
\end{figure}

Our idea is to detune the external laser slightly below the
transition $|g,0\rangle\to |-\rangle$ at $x=0$. If the laser
intensity is low enough, its only effect will be to produce an
AC--Stark shift for the level $|g,0\rangle$, whose energy $E_0(x)$
will now depend on the position, as shown in Fig.\ \ref{jc_stark}.
Thus, if the atom is in the level $|g,0\rangle$, it will
experience a trapping force towards $x=0$, and therefore, it can
be trapped (as long as the corresponding potential supports bound
states). Note also, that since the atom is basically in the ground
state and no photon is present, all the dissipative mechanisms may
be drastically reduced.

In the following sections we will compute the performance of our
scheme. In the rest of this section we will use very simple
estimates to characterize the trapping potential and the
corresponding time scales.

Denoting by $\Omega$ the Rabi frequency of the external laser, and
by $\Delta$ its corresponding detuning with respect to the
$|g\rangle\to |e\rangle$ transition ($\Delta<0$), we have that the
regime of validity of our analysis will be
 \be\label{approx}
 \Omega \ll |\Delta+g_0| \ll g_0.
 \ee
In this case, the depth of the trapping potential $V_0$ will be
approximately equal to the AC--Stark shift of the level
$|g,0\rangle$ due to its coupling to $|-\rangle$ at $x=0$, i.e.
 \be
 \label{V0}
 V_0 \simeq \frac{\Omega^2}{8\,|\Delta+g_0|}.
 \ee
On the other hand, losses will be due to the small contamination
of level $|g,0\rangle$ with level $|-\rangle$ given by the
off--resonant coupling. The population of this level will be of
the order of $\Omega^2/4|\Delta+g_0|^2$, and therefore the
lifetime of the state will be
 \be
 \label{tau}
 \tau\simeq
 \frac{4|\Delta+g_0|^2}{\Omega^2}\min(\Gamma^{-1},\kappa^{-1}).
 \ee
Equations (\ref{V0}) and (\ref{tau}) indicate that the lifetime
can be made arbitrarily big at the expense of reducing the
potential depth.

In three dimensions, one can easily estimate the condition for a
potential to possess a bound state. It is given by \cite{Galindo}
$2mV_0L^2/\hbar^2 \agt 1$, where $L$ is the cavity length, $m$ is
the atomic mass, and we have included $\hbar$ to make the
dimensions more explicit. We can rewrite this as
$V_0(L/\lambda)^2\agt E_R$, where $\lambda$ is an optical
transition wavelength and $E_R$ the corresponding energy of one
photon recoil. Since $L\agt\lambda$ in all cases we see that by
taking $V_0>E_R$ we will always have an atomic bound state. Note
that in a one dimensional set--up there is always a bound state
for any value of $V_0$ \cite{Galindo}.

So far, we have shown that it is possible to have atoms trapped in
the cavity with basically zero photons and in the atomic ground
state. However, the trapping potential may become very weak. Thus,
in order to trap atoms it will be required that they move very
slowly in the cavity in the state $|g,0\rangle$ and then, when
they are close to $x=0$, the external field is turned on. Let us
estimate what will be, in this case, the lifetime of the trapped
state. We will assume that we have Rb atoms and the kinetic energy
is of the order of one optical recoil ($E_R=\hbar^2k^2/2m$, where
$k$ is the optical wavevector). Thus, we take $E_R=4$kHz
$<V_0=10$kHZ. Let us analyze separately the optical and microwave
regimes.

%estimation of trapping time
%
For the optical regime we take the parameters from
\cite{Rempe00trap}. There the $5^2S_{1/2}F=3 \to 5^2P_{3/2}F=4$
transition of $^{85}{\rm Rb}$ with a frequency of
$3.8\times10^{14}{\rm Hz}$ was used. The maximal coupling between
cavity and atom is $g_0\approx 16\times2\pi\,{\rm MHz}$, the
cavity loss rate is $\kappa\approx 1.4\times2\pi\,{\rm MHz}$ and
the spontaneous decay rate is $\Gamma\approx 3\times2\pi\,{\rm
MHz}$. We estimate (\ref{tau}) a decay time of
$2.1\times10^{-5}\,{\rm s}$.
For the microwave regime we consider circular Rydberg states, so
we have \cite{Haroche91trap,Haroche96dec} $g_0\approx
67\times2\pi\,{\rm kHz}$, $\kappa\approx 1.6\times2\pi\,{\rm Hz}$
and $\Gamma\approx 1.6\times2\pi\,{\rm Hz}$, where $\Gamma$ is the
spontaneous decay rate of the Rydberg transition. We reach a life
time of $40\,{\rm s}$ (\ref{tau}).

These estimates look very promising. They will be optimized and
compared with numerical calculations in the following sections. On
the other hand, let us stress that we have calculated here the
time for a single loss event, since it will destroy the coherence
present in the atomic state. For the reference
\cite{Ritsch97,Ritsch98,Kimble01trap,Kimble99trap,Kimble00trap,
Rempe00trap,Rempe00trap_long}, in the optical experiments in which
the atom is trapped in a cavity both the effective decay rate and
the potential depth seen by the atom scale proportionally to the
population of the excited level. However, in our case the
potential depth (\ref{V0}) scales in a different way. Thus we
expect that our scheme will be useful under appropriate conditions
(small initial velocities). In the following sections we will also
analyze the trapping time if several loss events are allowed.

%%%%%%%%%%%%%%%%%%%%%%%%%%%%%%%%%%%%%%%%%%%%%%%%%%%%%%%%%%%%%
\section{Model}

In this section we will introduce in detail the model that
describes the situation we have in mind. In the first subsection
we will start with the full Hamiltonian characterizing the
atom--cavity interaction and perform some approximations in order
to derive the estimates given in the previous section. Then we
will introduce the decay mechanisms in this picture.

\subsection{Hamiltonian dynamics}

The Hamiltonian describing the dynamics of the atom and the cavity
mode can be written as follows
\begin{eqnarray}\label{Ham1}
H &=& \frac{p^2}{2m} + \omega_0 (a^\dagger a+ \frac{1}{2}\sigma_z)
+ g(x) (\sigma^+ a+a^\dagger \sigma^-) \nonumber\\&&+\,
\frac{\Omega}{2}(\sigma^+
 e^{-i\omega_L t} + \sigma^- e^{i\omega_L t}).
\end{eqnarray}
Here, $\omega_L$ and $\omega_0$ are the laser and atomic
transition frequency, respectively, $\Omega$ is the Rabi frequency
and $g(x)$ the position dependent coupling constant between the
cavity mode and the atom. Note that we have not included the
position dependence of the laser plane wave to make more explicit
the fact that the laser exerts no force on the atom (in any case,
since this laser only gives rise to AC-Stark shift, its position
dependence will cancel out).

In order to make the analysis simpler, we will project our system
in the subspace spanned by the states
$\{|g,0\rangle,|e,0\rangle,|g,1\rangle\}$. In any case, the reader
can easily verify that the population of all other levels will be
much smaller than the last two, which will be scarcely populated.
The Hamiltonian (\ref{Ham1}) in this subspace can be rewritten as
$H=p^2/2m + H'(x)$, and this last can be diagonalized exactly in
the rotating frame. Instead of doing that, we calculate the
eigenstates and eigenvalues of $H'(x)$ in lowest order
perturbation theory with respect to $\Omega$, which is assumed to
be small with respect to $|\Delta+g(x)|$ for all values of $x$
(see Fig.\ \ref{jc_stark}), where $\Delta=\omega_L-\omega_0$. We
obtain
\begin{eqnarray} \label{states}
|\Psi_0\> &=& |g,0\> + \frac{\Omega/2}{\Delta^2-g(x)^2}~\left(g(x)|g,1\> + \Delta |e,0\>\right) \nonumber\\
|\Psi_1\> &=& \frac{1}{\sqrt{2}}~\left( |g,1\> - |e,0\> - \frac{\Omega/2}{\Delta+g(x)} |g,0\>\right) \nonumber\\
|\Psi_2\> &=& \frac{1}{\sqrt{2}}~\left( |g,1\> + |e,0\> -
                                 \frac{\Omega/2}{\Delta-g(x)} |g,0\> \right)
\end{eqnarray}
and the corresponding eigenvalues
\begin{eqnarray} \label{values}
\lambda_0(x) &=& \frac{\Delta}{2} + \frac{\Omega^2}{8}~\left( \frac{1}{\Delta+g(x)} + \frac{1}{\Delta-g(x)} \right) \nonumber\\
\lambda_1(x) &=& -\frac{\Delta}{2} - g(x) - \frac{\Omega^2}{8(\Delta+g(x))} \nonumber\\
\lambda_2(x) &=& -\frac{\Delta}{2} + g(x) -
\frac{\Omega^2}{8(\Delta+g(x))}.
\end{eqnarray}

As we see, the ground state is basically $|g,0\rangle$ with a
vanishing contribution of levels $|g,1\rangle$ and $|e,0\rangle$
in the limit $\Omega\ll |\Delta+g(x)|$. However, it acquires a
position--dependent shift in its energy. The two terms in the
shift come from the AC--Stark shifts due to $|-\rangle$ and
$|+\rangle$, respectively, which, with the chosen detuning, do not
compensate each other. The shift is maximal at $x=0$. To obtain
the potential depth $V_0$ we have to subtract the shift at
$g(x)=0$. This gives
\begin{equation} \label{V0_exact}
V_0 = -\frac{\Omega^2 g_0^2}{4\Delta \left(\Delta^2-g_0^2\right)}.
\end{equation}
In the limit (\ref{approx}) we have $\Delta\approx -g_0$. If we
plug this into (\ref{V0_exact}) we obtain (\ref{V0}).

Starting out with $\Omega=0$, if the atom is initially in
$|g,0\rangle$ and has a small velocity near $x=0$, and we turn the
laser on, it will basically remain in the eigenstate
$|\Psi_0\rangle$, and therefore will experience the potential
$\lambda_0(x)$. Note that for this picture to be valid, we need
the kinetic energy of the atom to be smaller than $|\Delta+g_0|$
since in this case we can adiabatically eliminate the levels
$|g,1\rangle$ and $|e,0\rangle$ and obtain the effective
Hamiltonian
 \be\label{Hadia}
 H_{ad} = \frac{p^2}{2m} + \lambda_0(x)|g,0\rangle\langle g,0|.
 \ee

\subsection{Dissipation}

We introduce a cavity decay rate $\kappa$ and a spontaneous decay
rate $\Gamma$ for the atom. To take both into account we use the
master equation that describes the time evolution of this open
quantum systems. The state of the system, which is now in general
a mixed one, is given by a density matrix $\rho$. For our system
we obtain
 \be\label{master}
  \dot \rho =-i [H,\rho] + ({\cal L}_{\rm cav} + {\cal L}_{\rm at})
  \rho.
 \ee
Here,
 \be\label{Lcav}
 {\cal L}_{\rm cav}\rho = \kappa \left(2\,a\,\rho\,a^{\dagger}
                 -a^{\dagger}a\,\rho-\rho\,a^{\dagger}a \right)
 \ee
describes cavity damping, whereas
 \begin{eqnarray}\label{Latom}
 {\cal L}_{\rm at}\rho &=& \Gamma
 \left(2\,\int_{-1}^{1}\,N(u)~\sigma^-\,e^{-iux}\,\rho\,e^{iux}\,\sigma^+{\rm d}u
 \right.\nonumber\\&&
 \left. -\sigma^+ \sigma^-\,\rho-\rho\,\sigma^+ \sigma^-\right)
 \end{eqnarray}
describes spontaneous emission. The first term in this expression
accounts for the photon recoil experienced by the atom after
photon emission. We have included here a one dimensional version,
since in our numerical calculations we have investigated this
case.

%Discussion of jump formalism as is, including references...
%
To simulate a single trajectory we use the {\em Quantum Jump
Approach} \cite{Hegerfeldt92,Dalibard,Carmichael93,Dum92}.
Therefore one defines an effective non-Hermitian Hamiltonian
$H_{\rm eff}$ which describes the time evolution of the system
under the condition that no emission takes place. The master
equation can than be written in the form
\begin{eqnarray} \label{mastereff}
\dot{\rho} &=& -i\left[\left(H_{\rm eff}+
                     \frac{p^2}{2 m}\right),\rho\right] \nonumber\\
             & & +\,2\,\kappa\,a\,\rho\,a^{\dagger}
                 +2\,\Gamma\,\int_{-1}^{1}\,N(u)~\sigma^-\,e^{-iux}\,\rho\,e^{iux}\,\sigma^+{\rm d}u.\nonumber\\
\end{eqnarray}
The decay rates contribute to the effective time evolution as
damping terms. Therefore the norm of the state decreases. This
means that the probability to find no photon in the time interval
$(0,t)$ decreases with time "$t$".

Dissipation occurs in our model due to the small contamination of
level $|\Psi_0\rangle$ with the states $|g,1\rangle$ and
$|e,0\rangle$, which in turn decay due to cavity damping and
spontaneous emission, respectively. In order to determine the
effective decay rate (or jump time) we take the sum over the
probabilities for the excited states $|g,1\>$ and $|e,0\>$ in
$|\Psi_0\>$ weighted with the cavity decay rate $\kappa$ and the
spontaneous emission rate $\Gamma$. For the coupling constant we
assume that the atom is in the center of the cavity [$g(x)=g_0$].
We obtain
\begin{eqnarray} \label{gammaeff}
\Gamma_{\rm eff} & = & \kappa \cdot
\left|\frac{g_0\Omega/2}{\Delta^2-g_0^2}\right|^2 +
                   \Gamma \cdot \left|\frac{\Delta\Omega/2}{\Delta^2-g_0^2}\right|^2 \nonumber\\
             & = &
             \frac{\Omega^2\left(\kappa~g_0^2+\Gamma~\Delta^2\right)}{4\left(\Delta^2-g_0^2\right)^2}.
\end{eqnarray}
This gives an effective decay time of
\begin{equation} \label{teff}
\tau_{\rm eff} = \frac{1}{\Gamma_{\rm eff}}
               = \frac{4\left(\Delta^2-g_0^2\right)^2}{\Omega^2\left(\kappa~g_0^2+\Gamma~\Delta^2\right)}.
\end{equation}
For the estimation of the life time in Eq. (\ref{tau}) we
neglected the contribution of the upper dressed level $|+\>$. If
we consider this and the approximation $\Delta\approx -g_0$ and
plug it with $\kappa=\Gamma=max(\kappa,\Gamma)$ into (\ref{teff})
we end up with the expression (\ref{tau}).

\subsection{Discussion}

\begin{figure}
\includegraphics[width=8cm]{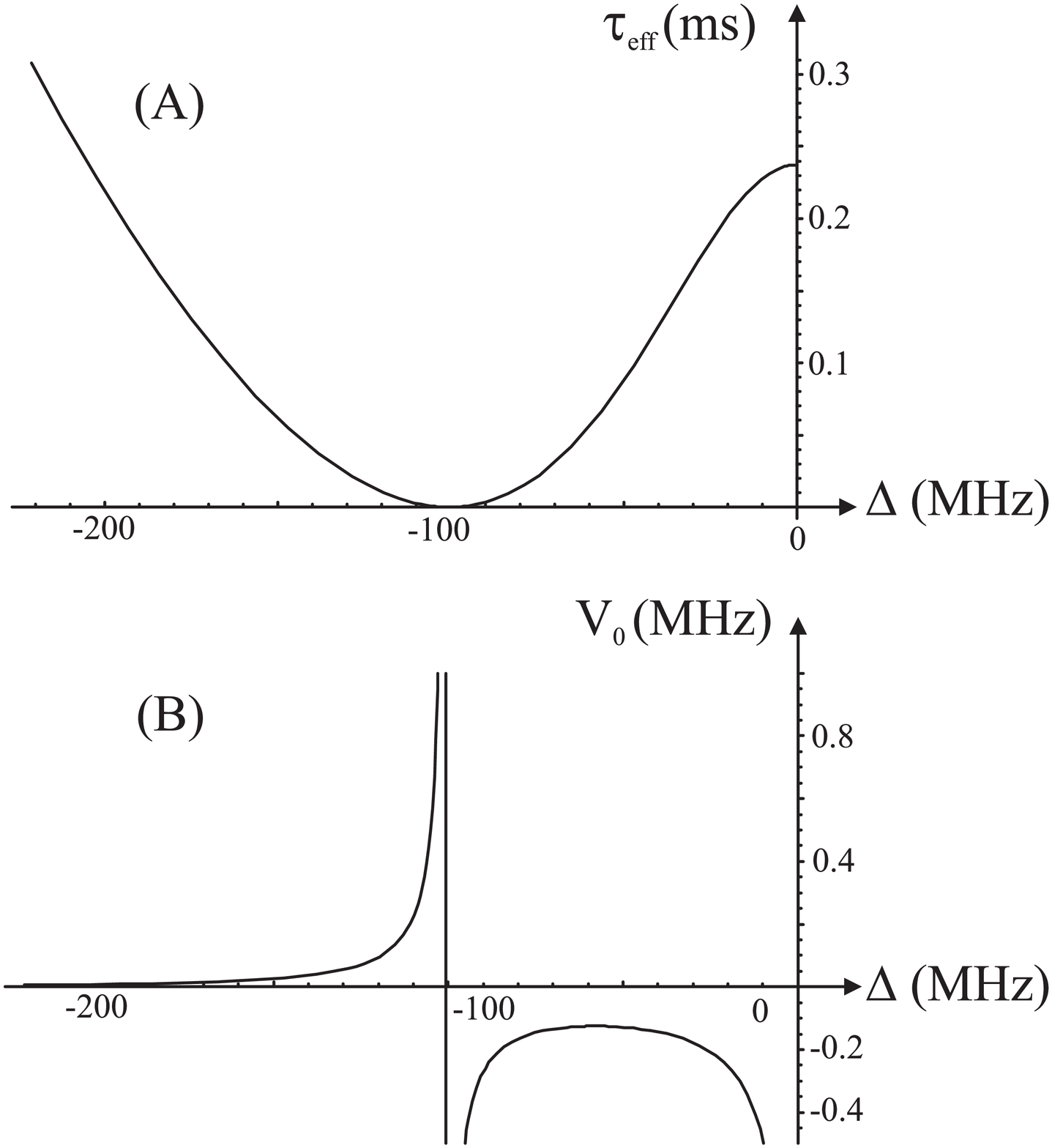}
\caption{Effective decay time $\tau_{\rm eff}$ (A) and potential
depth $V_0$ (B) versus laser detuning $\Delta=\omega_L-\omega_0$.
For the Rabi frequency of the laser we took
$\Omega=0.70\times2\pi{\rm MHz}$. The coupling strength between
cavity and atom is $g_0= 16\times2\pi\,{\rm MHz}$, the cavity loss
rate $\kappa= 1.4\times2\pi\,{\rm MHz}$ and the spontaneous decay
rate $\Gamma= 3\times2\pi\,{\rm MHz}$.} \label{teff_V0_delta}
\end{figure}

\begin{figure}
\includegraphics[width=8cm]{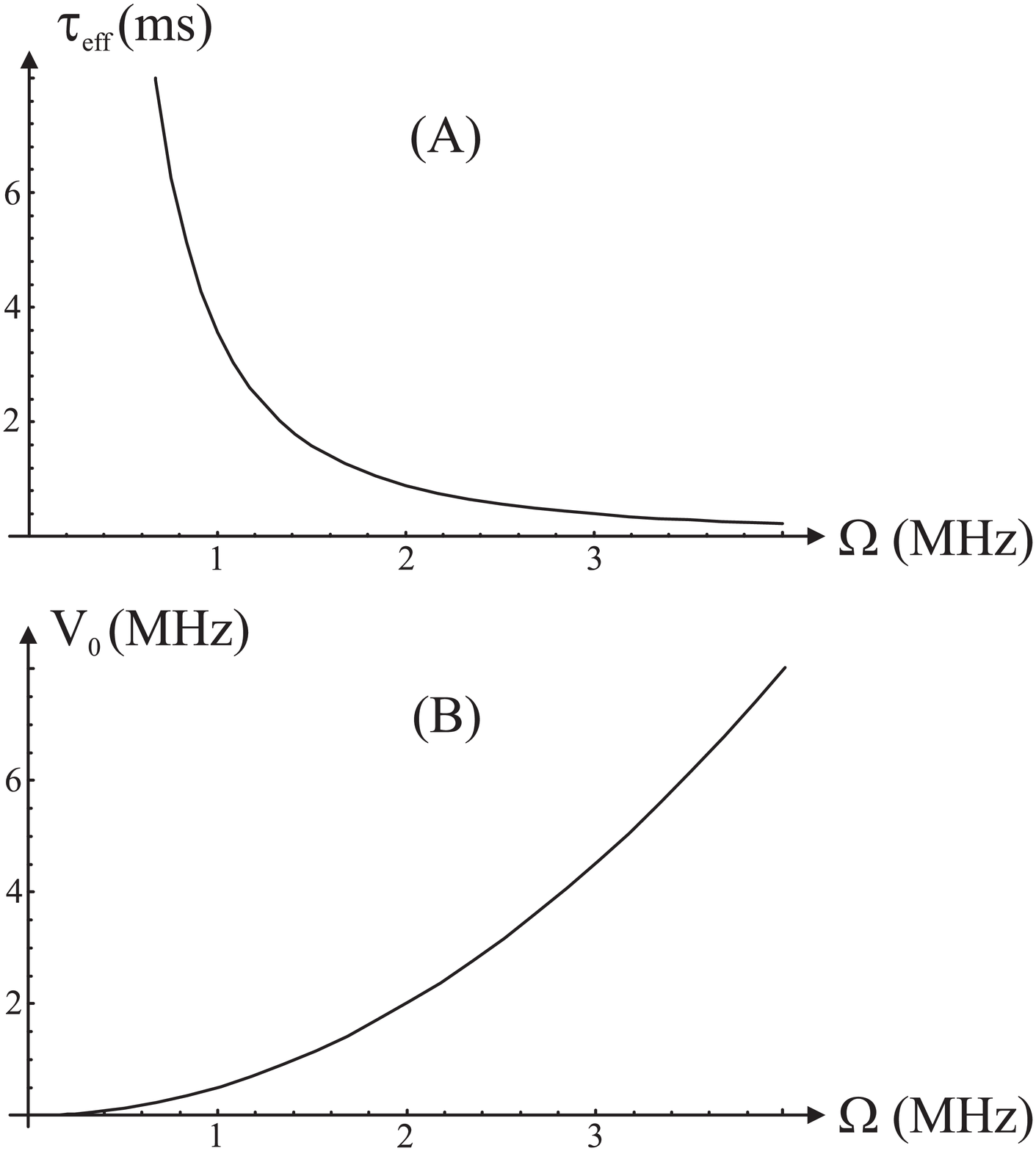}
\caption{Effective decay time $\tau_{\rm eff}$ (A) and potential
depth $V_0$ (B) versus Rabi frequency $\Omega$ of the laser. For
the laser detuning we took $|\Delta|=1.90\,g_0=30\times2\pi\,{\rm
MHz}$. The other parameters are the same as in Fig.\
\ref{teff_V0_delta}.} \label{teff_V0_omega}
\end{figure}

In Fig.\ \ref{teff_V0_delta} and Fig.\ \ref{teff_V0_omega} we have
plotted the potential depths $V_0$ and the effective life time
$\tau_{\rm eff}$ versus $\Delta$ and $\Omega$.
%
%Delta
%
From Fig.\ \ref{teff_V0_delta}(A) we see that in order to get a
long decay time it would be desirable to have $|\Delta|\gg|g_0|$.
In Fig.\ \ref{teff_V0_delta}(B) the region $-g_0<\Delta<0$ is not
of interest since there one obtains no attractive potential
($V_0<0$). One has to find a compromise between $\Delta$ close to
$-g_0$ in order to get a deep potential and $|\Delta|\gg|g_0|$ in
order to obtain a long decay time. This behavior is not surprising
because if the detuning is close to $-g_0$ the population in
$|-\>$ increases. This leads to a short decay time and a deep
potential.
%
%Omega
%
The same reasoning explains the plots in Fig.\ \ref{teff_V0_omega}
since the Rabi frequency $\Omega$ is a measure for the coupling
strength between the atomic transition and the laser.

%optimal parameters for fixed potential depth
%
For the parameters from \cite{Rempe00trap} and a potential depth
of $V_0=10\,{\rm kHz}$ the longest effective life time we can
achieve in the optical regime is $\tau_{\rm eff}=0.18\,{\rm ms}$.
The corresponding values for the laser parameters are
\begin{eqnarray} \label{laser_opt}
\Omega &=& 0.70\times2\pi\,{\rm MHz}~, \nonumber\\
|\Delta| &=& 1.90\,g_0=30\times2\pi\,{\rm MHz}.
\end{eqnarray}
In the microwave regime \cite{Haroche91trap,Haroche96dec} we get
$\tau_{\rm eff}= 1.26\,{\rm sec}$ with
\begin{eqnarray} \label{laser_mic}
\Omega &=& 54\times2\pi\,{\rm kHz}~, \nonumber\\
|\Delta| &=& 2.06\,g_0=0.14\times2\pi\,{\rm MHz}.
\end{eqnarray}
It is important to mention that since we used the expressions from
(\ref{V0_exact}) and (\ref{teff}) we are not in the limit
$\Delta\approx -g_0$ (\ref{approx}). This leads to a significantly
longer life time in the optical regime.

%%%%%%%%%%%%%%%%%%%%%%%%%%%%%%%%%%%%%%%%%%%%%%%%%%%%%%%%%%%%%
\section{Numerical results}

Here we investigate the behavior of the system numerically. For
the analytic results we made certain approximations. The
comparison with the numerical results will show if these
assumption are justified for realistic parameters. Furthermore we
will include spontaneous emission and photon recoil.

We denote the state of the system by $|\Phi\>$. For the simulation
we write it as $|\Phi\> = |\varphi_{g0}\> + |\varphi_{g1}\> +
|\varphi_{e0}\>$, where $|\varphi_i\> = |i\>\<i|\Phi\>$. We
consider only the contributions of $|g,0\>$, $|g,1\>$ and $|e,0\>$
since the population of the levels with two and more excitations
is negligible. As for the analytic estimations we restrict the
investigations to one dimension. The probability amplitudes for
the system being in the states $|g,0\>$, $|g,1\>$ and $|e,0\>$ at
position "$x$" are given by $\varphi_{g0}(x) =
\<x|\varphi_{g0}\>$, $\varphi_{g1}(x) = \<x|\varphi_{g1}\>$ and
$\varphi_{e0}(x) = \<x|\varphi_{e0}\>$.

In the first subsection we calculate the ground state of the
system with and without the assumptions made above. In the
following we include dissipation and compare the decay time with
the effective decay time $\tau_{\rm eff}$ we estimated in the last
section. Finally we consider spontaneous emissions and recoil and
investigate how long the atom remains in the cavity for different
parameters. Apart from one simulation with the parameters from the
analytic estimation we will only consider the optical regime in
this section.

\subsection{The ground state}

To obtain the ground state we apply the {\em imaginary time
evolution} \cite{Zoller99} to an arbitrary initial state until it
remains unchanged. Instead of the time evolution operator
$e^{-{\rm i} H \Delta t}$ one uses a modified operator $e^{-H
\Delta t}$. After a sufficient number of iterations this damps
away all states orthogonal to the one with the lowest eigenvalue,
which is the ground state of the system.

\begin{figure}[ht]
\includegraphics[width=8.5cm]{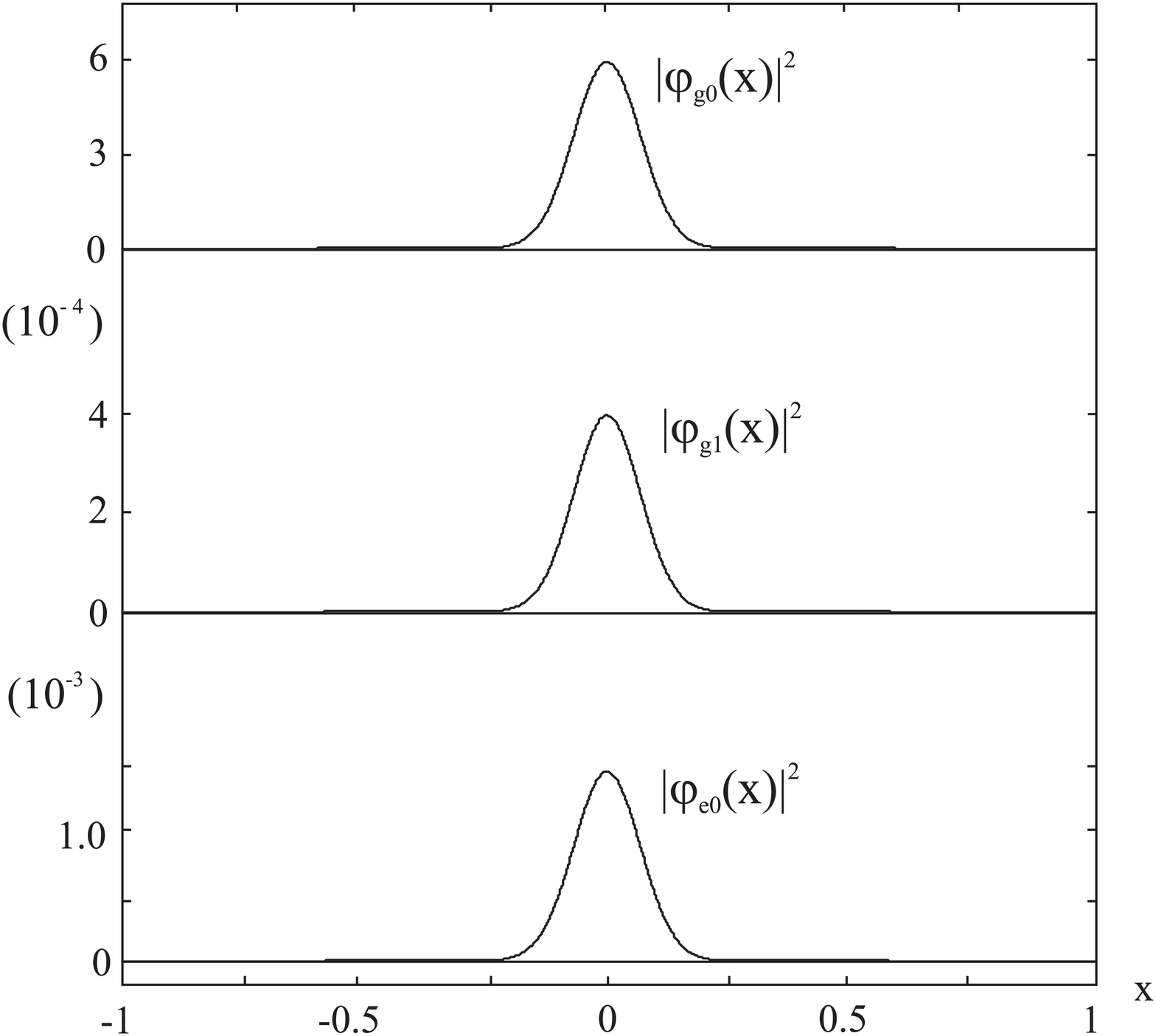}
\caption{Numerical simulation of the ground state. The plots show
$|\varphi_{g0}(x)|^2$, $|\varphi_{g1}(x)|^2$ and
$|\varphi_{e0}(x)|^2$ versus $x$. They satisfy the normalization
condition $\int\left(|\varphi_{g0}(x)|^2+|\varphi_{g1}(x)|^2+
|\varphi_{e0}(x)|^2\right)dx=1$. The plots are in units of
$\sigma$, which is the cavity width. For the laser detuning we
took $\Delta=1.90\,g_0=30\times2\pi\,{\rm MHz}$ and for the Rabi
frequency of the laser $\Omega=0.70\times2\pi{\rm MHz}$. The
coupling strength between cavity and atom is $g_0=
16\times2\pi\,{\rm MHz}$, the cavity loss rate $\kappa=
1.4\times2\pi\,{\rm MHz}$ and the spontaneous decay rate $\Gamma=
3\times2\pi\,{\rm MHz}$.} \label{ground}
\end{figure}

The Hamiltonian of the system is given in Eq. (\ref{Ham1}). We
denote its ground state by $|\Phi_0\>$. For $\Omega$ and $\Delta$
we took the values from (\ref{laser_opt}). According to the
analytic approximation they should give the maximal decay time
which is achievable for a potential depth of $10\,{\rm kHz}$. The
numerical simulation of the ground state leads to the probability
distribution shown in Fig.\ \ref{ground}.
The three plots show the population distribution of the three
internal states separately. The excited states are only very
weakly populated. The probability to find an atom in the center of
the cavity with the system being in state $|g,1\>$ or $|e,0\>$ is
three to four orders of magnitude smaller than to find it there
with the state $|g,0\>$. We also found that the atom is well
localized in the center of the cavity. At $0.1\,\sigma$, where
$\sigma$ is the width of the cavity, the probability to find the
atom is already reduced by more than $1/2$.

In order to valuate the approximations made for the analytic
estimation we calculated the ground state also using the
Hamiltonian from Eq. (\ref{Hadia}). We denote its ground state
solution by $|\xi_0\>$. We find
\begin{equation}
|\Phi_0\>\approx|g,0\>\otimes|\xi_0\>.
\end{equation}
This means that nearly all the population is in $|g,0\>$. So the
approximations in the analytical approach are justified and one
can trap an atom in a basically empty cavity.

\subsection{Dissipation and photon emissions}

In this subsection we include the coupling of the system to the
environment and as a consequence the spontaneous emission of
photons. First we will only consider the time evolution of the
system under the condition that no photon is emitted and compare
the decay time with the analytic estimation. Then we include also
spontaneous emissions and the recoil kick the atom experiences.

In the Quantum Jump Approach one describes the time evolution of
the system with an effective Hamiltonian as long as no photon is
emitted. The emissions which cause the system to jump in a
different state are described by reset operators. We obtain an
expression for the effective Hamiltonian by comparing Eqs.
(\ref{master}),(\ref{Lcav}) and (\ref{Latom}) with Eq.
(\ref{mastereff}). This gives
\begin{eqnarray} \label{haminteff}
H_{\rm eff} &=& \frac{p^2}{2m}-\frac{\Delta}{2}\,\left(|g,1\>\<g,1|+|e,0\>\<e,0|-|g,0\>\<g,0|\right) \nonumber\\
        & & +\,g(x)\,\left(|e,0\>\<g,1|+|g,1\>\<e,0|\right) \nonumber\\
        & & +\,\frac{\Omega}{2}\left(|g,0\>\<e,0|+|e,0\>\<g,0|\right) \nonumber\\
        & & -\,{\rm i}\kappa\,|g,1\>\<g,1|-{\rm
        i}\Gamma\,|e,0\>\<e,0|~,
\end{eqnarray}
where we used an interaction picture rotating with the laser
frequency $\omega_L$ and assumed that there is at most one
excitation in our system.

After preparing the system in the ground state $|\Phi_0\>$ using
the imaginary time evolution we simulated the time evolution with
the effective Hamiltonian $H_{\rm eff}$. This leads to a damping
of the state of the system. So the probability $|\Phi|^2$ that no
photon has been emitted also decreases. We compare the time after
which this probability has reached $1/e$ with the effective life
time $\tau_{\rm eff}$ we estimated analytically. For the
parameters from (\ref{laser_opt}) we obtained $\tau_{\rm
eff}=0.18\,{\rm ms}$, which agrees with the decay time from the
numerical simulation of $0.14\,{\rm ms}$.

%jumps
%
In the Quantum Jump Approach the jumps are described by reset
operators. We obtain them from the master equation
(\ref{mastereff}). For the spontaneous emission of the atom we get
\begin{equation} \label{jumpatom}
e^{-iux}\sqrt{2\Gamma}|g,0\>\<e,0|~,
\end{equation}
where $e^{-iux}$ describes the momentum shift "$u$" due to the
photon recoil. If the cavity emits a photon one has to apply
\begin{equation} \label{jumpcav}
\sqrt{2\kappa}~|g,0\>\<g,1|.
\end{equation}
In both cases the population of the excited level gets shifted to
the ground state. After that one has to normalize the wave
function.

In the following we will discuss the trapping time $\tau_{\rm
trap}$ of the atom. We define it as the time when the probability
to find the atom in the cavity ($|x|<\sigma$) is reduced to $0.5$.
The atom has an initial kinetic energy and gains a momentum kick
when it spontaneously emits a photon. When the motional energy is
bigger than the trapping potential the atom leaves the cavity. So
it is desirable to achieve a long decay time in order to get a low
photon emission rate. On the other hand a deeper potential
provides the possibility of a bound state for an atom which
experienced more recoil kicks. From our analytic estimations we
know that these demands contradict each other.
For the simulation we took the parameters from (\ref{laser_opt})
for a potential depth of $V_0=10 {\rm kHz}$ and the longest
corresponding effective decay time $\tau_{\rm eff}=0.18\,{\rm
ms}$. We applied the operator $e^{-iux}$ to the ground state which
we got from the imaginary time evolution in order to take into
account an initial kinetic energy. The momentum shift "u"
corresponds to the energy of one photon recoil $E_R=4\,{\rm kHz}$.
After a couple of spontaneous emissions the atom leaves the cavity
at $\tau_{\rm trap}=0.73\,{\rm ms}$.

\begin{figure}[ht]
\includegraphics[width=8cm]{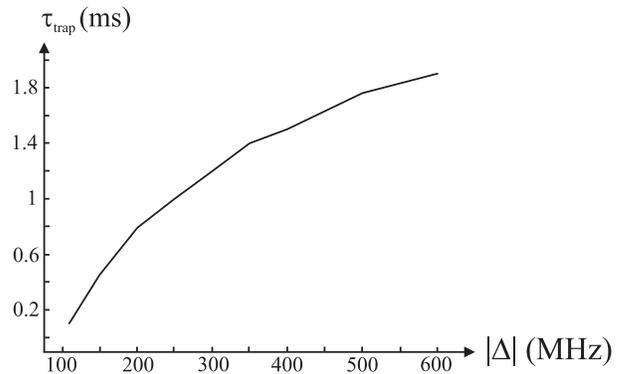}
\caption{Numerical results for the trapping time $\tau_{\rm trap}$
versus detuning $\Delta$ for a Rabi frequency
$\Omega=0.70\times2\pi{\rm MHz}$. The coupling strength between
cavity and atom is $g_0= 16\times2\pi\,{\rm MHz}$, the cavity loss
rate $\kappa= 1.4\times2\pi\,{\rm MHz}$ and the spontaneous decay
rate $\Gamma= 3\times2\pi\,{\rm MHz}$.} \label{t_delta}
\end{figure}

In order to achieve a longer trapping time we first varied the
detuning $\Delta$ and left the Rabi frequency
$\Omega=0.70\times2\pi{\rm MHz}$ unchanged. The result is shown in
Fig.\ \ref{t_delta}.
A larger detuning of the laser leads to a smaller potential depth
$V_0$ and a longer effective life time $\tau_{\rm eff}$. From
Fig.\ \ref{t_delta} we see that the longer life time has a bigger
influence on the trapping time since $\tau_{\rm trap}$ increases
with growing detuning. It is not surprising that the effective
life time has a crucial influence on the trapping time since it
determines how fast the motional energy of the atom grows.

\begin{figure}[ht]
\includegraphics[width=8cm]{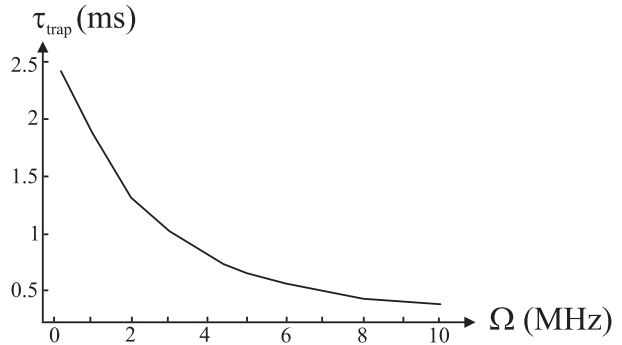}
\caption{Numerical results for the trapping time $\tau_{\rm trap}$
versus Rabi frequency $\Omega$ for a detuning
$|\Delta|=1.90\,g_0=30\times2\pi\,{\rm MHz}$. The coupling
strength between cavity and atom is $g_0= 16\times2\pi\,{\rm
MHz}$, the cavity loss rate $\kappa= 1.4\times2\pi\,{\rm MHz}$ and
the spontaneous decay rate $\Gamma= 3\times2\pi\,{\rm MHz}$.}
\label{t_omega}
\end{figure}

A larger Rabi frequency causes a smaller effective life time and a
deeper potential. So consistently we expect a decreasing trapping
time when we enlarge the Rabi frequency. This is confirmed by
Fig.\ \ref{t_omega}, where we plotted $\tau_{\rm trap}$ versus
$\Omega$ for a fixed detuning $|\Delta| = 1.90\,g_0 =
30\times2\pi\,{\rm MHz}$.
If we compare this plot with the plot in Fig.\
\ref{teff_V0_omega}(A) we ascertain a qualitative agreement. This
is again what we expect if we assume that the decisive variable
for the trapping time $\tau_{\rm trap}$ is the effective decay
time $\tau_{\rm eff}$.

The longest trapping times we can achieve in the optical regime
are of the order of $1\,{\rm ms}$. For the microwave regime we
took $g_0= 67\times2\pi\,{\rm kHz}$, $\kappa= 1.6\times2\pi\,{\rm
Hz}$ and $\Gamma= 1.6\times2\pi\,{\rm Hz}$. The trapping time we
obtained for the laser parameters from (\ref{laser_mic}) was
$\tau_{\rm trap}\approx 10\,{\rm s}$. The reason for the good
result are the very small decay rates for the Rydberg state and
the micro--cavity. Another important advantage over the optical
regime is that the recoil due to spontaneous emissions is
practically zero.

%%%%%%%%%%%%%%%%%%%%%%%%%%%%%%%%%%%%%%%%%%%%%%%%%%%%%%%%%%%%%
\section{Conclusions}

%it is possible
%how did we achieve it
%
We showed that it is possible to trap an atom in the vacuum field
of a high Q cavity. To do this we need a weak laser which couples
directly to the atom in the cavity. It induces a position
dependent AC Stark shift to the ground state of the cavity-atom
system. We use this energy shift as a trapping potential and as we
showed by an analytic estimation and a numerical simulation it is
deep enough to trap an atom with a realistic initial momentum.

%advantage/disadvantage
%
The advantage of this approach is the low effective decay rate due
to the little amount of excitation in the system. This requires to
cool the atom to a lower kinetic energy than the potential depth.
In order to obtain a long life time it would be good to have an
initial kinetic energy of the order of one photon recoil. This is
still difficult to achieve, even though it is possible to cool an
atom below one photon recoil with the method of velocity-selective
coherent population trapping \cite{Cohen} or Raman-cooling
\cite{Chu}. Another possibility would be a cavity assisted cooling
method \cite{Ritsch97,Ritsch98}.

%comparison with others
%
The trapping time we can achieve in the optical regime with our
approach is of the same order or even lower as observed already in
experiments \cite{Kimble99trap,Kimble00trap,
Rempe00trap,Rempe00trap_long}. The benefit of this method is that
the time after which the first jump occurs is longer because there
is only very little excitation in the system. As mentioned before
this decay time is very important for any kind of quantum
information application since the jump destroys the coherence in
the atomic state. In the microwave regime the trapping times can
be much longer.

\narrowtext {\em Acknowledgment.} We would like to thank G. Rempe,
G. Giedke, J. J. Garcia-Ripoll and B. Kraus for interesting and
stimulating discussions.

\end{document}